\documentclass[prper,aps,twocolumn,groupedaddress,floats,showpacs,final,superscriptaddress]{revtex4-1}
\usepackage{graphicx}
\usepackage{dcolumn}
\usepackage{bm}
\usepackage{amssymb}
\usepackage{amsmath}
\usepackage{ulem}
\usepackage[english]{babel}
\input{epsf}

\begin{document}

\title{Orbital momentum excitation by interband optical transitions in a 2D system illuminated by twisted light}

\author{L.S. Braginsky}
\affiliation{Rzhanov Institute of Semiconductor Physics, \\
Siberian Branch of the Russian Academy of Sciences, Novosibirsk 630090, Russia}
\affiliation{Novosibirsk State University, Novosibirsk 630090, Russia}

\author{M.M. Mahmoodian}
\affiliation{Rzhanov Institute of Semiconductor Physics, \\
Siberian Branch of the Russian Academy of Sciences, Novosibirsk 630090, Russia}
\affiliation{Novosibirsk State University, Novosibirsk 630090, Russia}

\author{L.I. Magarill}
\affiliation{Rzhanov Institute of Semiconductor Physics, \\
Siberian Branch of the Russian Academy of Sciences, Novosibirsk 630090, Russia}
\affiliation{Novosibirsk State University, Novosibirsk 630090, Russia}

\author{M.V. Entin}
\affiliation{Rzhanov Institute of Semiconductor Physics, \\
Siberian Branch of the Russian Academy of Sciences, Novosibirsk 630090, Russia}

\begin{abstract}
Illumination of a two-dimensional system by a twisted light beam is considered in order to find specific effects caused by twisting. Direct interband transitions between valence and conduction bands are supposed. The generation rates of electron orbital momentum are found. A kinetic equation for an orbital momentum distribution function is formulated and solved. The mean electron orbital momentum is found.
\end{abstract}

\maketitle

\section{Introduction}
The twisted light states (states with the orbital angular momentum (OAM)) attracted attention having been growing since the publication by Allen et al in 1992 \cite{Allen}. This research direction found an essential application in the atomic and high-energy physics due to its additional information for scattering processes \cite{Yao,Torres,Andrews,Mak}.

The OAM led to a number of discoveries both at the microscopic and macroscopic level. It has proven useful for applications in micromanipulation, imaging and communication systems \cite{Wisniewski}.

In particular, the Compton backscattering of twisted laser photons on the ultrarelativistic electrons is a way to produce the high-energy twisted photons and to the pair production of nuclei in previously unexplored experimental regimes~\cite{serb1,serb2,Knyazev Serbo}.

The orbital contribution provides a fundamentally new degree of freedom with fascinating and wide-spread applications. The orbital angular momentum arises as a consequence of the spatial distribution of optical field intensity and phase. Twisted photons can be used to encode information beyond one bit per single photon. This ability offers a great potential for quantum information tasks, as well as for the investigation of fundamental issues.

Twisted photon waveguides ensure a new direction in photonics allowing a new degree of freedom of light with the orbital angular momentum in information processes and physics of low dimensions.

The photoexcitation by twisted light depends on the non diagonal components in the translational momentum of scattering matrices are hidden studying scattering probabilities. To this extent, twisted light-induced effects have an analogy to the displacement photogalvanic effect \cite{Belinicher,Bel-Ivch}, which is also caused by the non-diagonal scattering matrix elements, in other words, by the wave package shift during the scattering \cite{we}.

There are few papers devoted to twisted light-induced effects in bulk and 2D  semiconductors. The interband transitions and quantum kinetics induced by twisted light in bulk semiconductors were studied theoretically in \cite{aa,bb} emphasizing pulse processes. The spin polarization of photoelectrons excited by twisted photons in 2D semiconductor systems was considered in \cite{c}. Unlike \cite{c}, here we focus on the study of the orbital momentum of photoelectrons caused by twisted light in a direct-band semiconductor with a zinc blend lattice, such as GaAs. However, the orbital momentum transfer was not studied either.

The purpose of the present paper is the study of the orbital momentum generation due to the twisted light causing the interband transition between the central extrema of a zinc blend-like semiconductor.

We study the orbital momentum excitation in 2D systems. In general, the orbital momentum is connected with the spin via the spin-orbit interaction and can be transferred to this degree of freedom. However, we shall consider the situation when the spin and orbital momentum are separate \cite{Dyak-Koch,Gan}. In this case, one can expect that the orbital momentum transferred to electrons will survive up to the recombination moment.

The orbital momentum can be observed via luminescence. Besides, the orbital momentum, being transferred to charged carriers, produces a principally observable magnetic momentum of a system.

For our purposes, we need the local action of twisted photons. This can be achieved by the application of twisted photon waveguides (TPW). Different ways to elaborate the twisted photon waveguides were offered. Two hypothetic systems of such kind are presented in Figures \ref{fig1} and \ref{fig2}. In what follows we will base on the simplest cylindrical TPW in which the orbital momentum (OM) mode can propagate.

The paper is organized as follows. First, we shall consider the field in the circular multimode dielectric optical guide. Then, the basic Hamiltonian, which determines the interaction of 2D carriers with the interband electromagnetic field will be presented. Third, axially symmetric electron states will be found. Then, the transition amplitudes will be obtained. After that, we shall consider the stationary orbital momentum caused by the stationary illumination in the 2D plane. Then, the problem of repumping between the orbital and spin momentum will be discussed. The last section of the paper is devoted to the discussion of the obtained results.

\section*{Problem formulation}
We shall study the optical transitions in a III-V direct band-gap semiconductor with a zinc blende crystal structure like GaAs caused by the electromagnetic field with a fixed orbital momentum projection in and outside the circular symmetric dielectric light guide. Our purpose is pumping the orbital momentum to electrons. Hence, we shall study the axially symmetric system where the orbital momentum projection is a conserving quantity.

Namely, we shall consider the circular waveguide normally connected to the 2D plane. The light frequency is supposed to be close to the edge of fundamental optical transitions. In this case, pumped electrons and holes have a small excitation energy and can conserve the orbital momentum long enough.

\section*{Twisted photons}
In the literature different lightguide sources of the twisted light were discussed \cite{Torres,pag,twist2,twist3,twist4,twist5,twist6,twist7,twist8}. To obtain the photoillumination sensitive to twisting, we assume using a local source. For example, such a source can serve as cylindrical light guide. This system has a rotation symmetry, with respect to axis $z$. The photon states with momenta $\pm m$ are degenerate. If we excite a photon in this state in one waveguide end, it will propagate along the waveguide without changes.

\begin{figure}[ht]
\centerline{\epsfysize=5cm\epsfbox{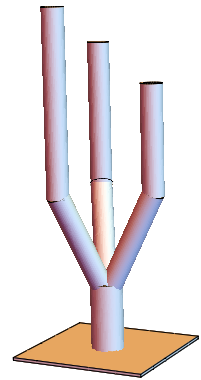}}
\caption{The conjuncture of 3 cylindrical light guides producing the azimuthal dependence of electromagnetic field. The phase shifts between lightguides $2\pi/3$ are provided by their length difference, produces the azimuthal dependence of the electromagnetic field in the final lightguide. This lightguide vertically illuminates the 2D system, thus, causing interband transitions.}\label{fig1}
\end{figure}

The selective excitation of state $m$ separately from state $-m$ can be done, for example, using of 3 or more waveguides of different lengths, connecting to a single lightguide. The length difference of the light guides and the consequent phase difference yield the continuous phase shift in the final lightguide (Figure \ref{fig1}).

For example, the conjunction of 3 waveguides with $2\pi/3$ phases shifted into a single waveguide would yield the $m=\pm 1$ wave propagating in a single waveguide. We suppose that these modes have real wavevectors.  This allows us to consider the wave with single angular momentum $m$.

\begin{figure}[ht]
\centerline{\epsfysize=1.5cm\epsfbox{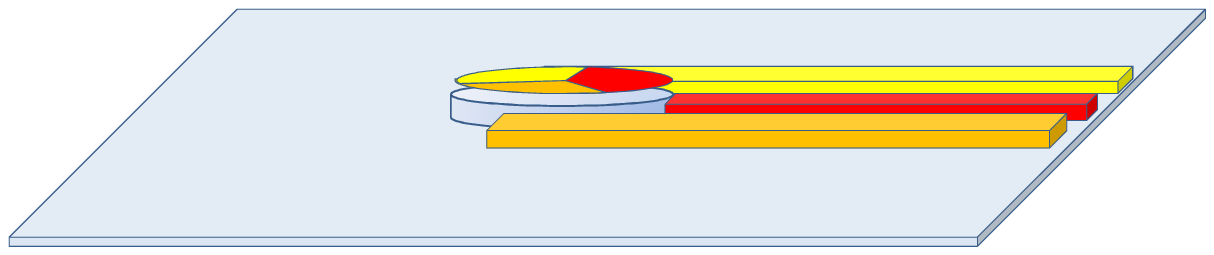}}
\caption{Planar lightguides light transport to the optical cavity from an emitting diode laser. The difference in the light guide lengths produces the phase shifts $2\pi/3$ between light guides. That, in turn, excites the twisting wave in the cylindrical cavity. The semiconductor quantum dot is located coaxially under the cavity.}\label{fig2}
\end{figure}

Another hypothetical variant is shown in Figure \ref{fig2}. Three planar light guides come out of a diode laser. They have the $2\pi/3$ relative phase shifts and combine to illuminate a round cavity under which the round quantum dot is coaxially located.

\section*{Field in the waveguide}
Consider a cylindrical dielectric  optical fiber with a step-like radial dependence of permittivity $\epsilon(r)$:
\begin{equation}
\epsilon(r)=\left\{
\begin{array}{ll}
    \varepsilon, & r\leq a, \\
    \varepsilon_1, & r> a,
\end{array}\right.
\end{equation}
where $\varepsilon>\varepsilon_1$. Looking for the solution of the Maxwell equations in the form ${\bf E}=({\cal E}_r,{\cal E}_\varphi,{\cal E}_z)$, ${\bf H}=({\cal H}_r,{\cal H}_\varphi,{\cal H}_z)$, ${\cal E}_z={\cal E}_z(r) e^{i(m\varphi+qz-\omega t)}$, ${\cal H}_z={\cal H}_z(r) e^{i(m\varphi+qz-\omega t)}$, we find
\begin{eqnarray}\label{20}
  &&{\cal E}_r=\frac{1}{\frac{\epsilon\omega^2}{c^2}-q^2}\left(iq\frac{\partial {\cal E}_z}{\partial r}-\frac{m\omega}{cr} {\cal H}_z\right), \nonumber\\
  &&{\cal E}_\varphi=-\frac{1}{\frac{\epsilon\omega^2}{c^2}-q^2}\left(\frac{mq}{r} {\cal E}_z+\frac{i\omega}{c}\frac{\partial {\cal H}_z}{\partial r}\right),\\
  &&{\cal H}_r=\frac{1}{\frac{\epsilon\omega^2}{c^2}-q^2}\left(\frac{\epsilon m\omega}{cr} {\cal E}_z+iq\frac{\partial {\cal H}_z}{\partial r}\right), \nonumber\\
  &&{\cal H}_\varphi=\frac{1}{\frac{\epsilon\omega^2}{c^2}-q^2}\left(\frac{i\epsilon\omega}{c}\frac{\partial {\cal E}_z}{\partial r}-\frac{m q}{r} {\cal H}_z\right). \nonumber
\end{eqnarray}
Here $e$ is the electron charge and $c$ is the light velocity. The boundary conditions at the surface $r=a$ allow us to consider two types of the twisted ($m\neq 0$) waveguide modes: the ${\cal E}_{\bm \tau}$ mode, where ${\cal E}_r=0$ and the ${\cal H}_{\bm \tau}$ mode, where ${\cal H}_r=0$. For the ${\cal E}_{\bm \tau}$ mode the electric field components are
\begin{eqnarray}\label{80}
  &&{\cal E}_r=0,\nonumber \\
  &&{\cal E}_\varphi=-\frac{qr}{m}{\cal E}_z, \nonumber\\
  &&{\cal H}_r=\left(\frac{mc}{\omega r}+\frac{q^2cr}{m\omega}\right){\cal E}_z, \\
  &&{\cal H}_\varphi=\frac{ic}{\omega}\frac{\partial {\cal E}_z}{\partial r},\nonumber\\
  &&{\cal H}_z=\frac{iqcr}{m\omega}\frac{\partial {\cal E}_z}{\partial r},\nonumber\\
  &&[{\cal E}_z]=0,\;\;\;\left[\frac{\partial {\cal E}_z}{\partial r}\right]=0\nonumber
\end{eqnarray}
and the ${\cal H}_\tau$ mode, where
\begin{eqnarray}\label{81}
  &&{\cal H}_r=0,\nonumber \\
  &&{\cal H}_\varphi=-\frac{qr}{m}{\cal H}_z,\nonumber\\
  &&{\cal E}_r=-\left(\frac{mc}{\epsilon\omega r}+\frac{q^2cr}{\epsilon m\omega}\right){\cal H}_z, \\
  &&{\cal E}_\varphi=-\frac{ic}{\omega\epsilon}\frac{\partial {\cal H}_z}{\partial r},\nonumber \\
  &&{\cal E}_z=-\frac{iqcr}{\epsilon\omega m}\frac{\partial {\cal H}_z}{\partial r},\nonumber \\
  &&[{\cal H}_z]=0,\;\;\;\left[\frac{1}{\epsilon}\frac{\partial {\cal H}_z}{\partial r}\right]=0.\nonumber
\end{eqnarray}

The square brackets in Eqs.~(\ref{80}) and (\ref{81}) denote the appropriate values discontinuity at the boundary $r=a$. For the ${\cal E}_\tau$ mode we write:
\begin{eqnarray}\label{50}
&&{\cal E}_z(r,\varphi,z)={\cal E}_0e^{im\varphi+iqz-i\omega t}\left\{
   \begin{array}{cc}
        J_m(\kappa r), & r\leq a, \\
        BK_m(\kappa_1r), & r> a.
   \end{array}\right.
\end{eqnarray}
Here $J_m(x)$ is the Bessel function of the first kind, $K_m(x)$ is the modified Bessel function of the second kind, $\kappa^2=\varepsilon\omega^2/c^2-q^2$ and $\kappa_1^2=-\varepsilon_1\omega^2/c^2+q^2$. The boundary conditions yield
\[
B=\frac{J_m(\kappa a)}{K_m(\kappa_1a)},
\]
and the dispersion law is
\begin{equation}\label{53}
  \frac{\kappa J_m'(\kappa a)}{J_m(\kappa a)}=\frac{\kappa_1K_m'(\kappa_1a)}{K_m(\kappa_1a)}.
\end{equation}

For the ${\cal H}_\tau$ mode,
\begin{eqnarray}\label{55}
&&{\cal H}_z(r,\varphi,z)={\cal H}_0e^{im\varphi+iqz-i\omega t}\left\{
  \begin{array}{cc}
       J_m(\kappa r), & r\leq a, \\
       DK_m(\kappa_1r), & r> a,
   \end{array}\right.
\end{eqnarray}
\[
D=\frac{J_m(\kappa a)}{K_m(\kappa_1a)},
\]
and the dispersion law
\begin{equation}\label{55}
  \frac{\kappa J_m'(\kappa a)}{\varepsilon J_m(\kappa a)}=\frac{\kappa_1K_m'(\kappa_1a)}{\varepsilon_1K_m(\kappa_1a)}.
\end{equation}

Note that $\kappa_1 \to 0$ at the propagation threshold, and that means an infinite increase of the effective transversal wave size. Therefore, one can neglect the field inside the lightguide and use only its external value at a relatively small $\kappa_1$.

To calculate the orbital momentum generation rate, we have to normalize the field for the flow of unite quanta $I$, which is expressed via the Poynting vector. Instead, we find $I$ by multiplying the energy density per unite lightguide length by the group velocity $d\omega/dq$ and division by the quantum energy $\omega(q)$:
$$I=\frac{d(\ln\omega)}{dq}\int_0^\infty rdr\epsilon(r){\cal E}^2(r)/2.$$
Here and below we assume $\hbar=1$.

\section*{Axial electron states in the 2D system based on III-V direct band-gap semiconductor}
We shall consider the 2D electron system based on the III-V direct band gap semiconductor with a zinc blende crystal structure. The fundamental interband transitions are assumed. The system spectrum is formed from the parabolic conduction-band and valence-band subbands splitted due to the transversal quantization.

We use the basis of states \cite{Ivch}
\begin{eqnarray}\label{basis}
|S \alpha \rangle,~~ | S \beta\rangle,\nonumber\\
  | 3/2,3/2 \rangle=-\frac{1}{\sqrt{2}}(X+ iY)|\alpha\rangle,\nonumber\\
    | 3/2,1/2 \rangle=-\frac{1}{\sqrt{6}}((X+iY)|\beta\rangle-2Z|\alpha\rangle),\nonumber\\
      | 3/2,-1/2 \rangle=\frac{1}{\sqrt{6}}((X-iY)|\beta\rangle+2Z|\alpha\rangle),\\
        | 3/2,-3/2 \rangle=\frac{1}{\sqrt{2}}(X-iY)|\beta\rangle,\nonumber\\
          | 1/2,1/2 \rangle=-\frac{1}{\sqrt{3}}((X+iY)|\beta\rangle+Z|\alpha\rangle),\nonumber\\
            | 1/2,-1/2 \rangle=-\frac{1}{\sqrt{3}}((X-iY)|\alpha\rangle-Z|\beta\rangle).\nonumber
\end{eqnarray}
The symbols $\alpha$ and $\beta$ stand for spinors $(1,0)$ and $(0,1)$, respectively. States $|S\rangle$ and $|S\beta\rangle$ are singlet states from the conduction band bottom; states $X,Y,$ and $Z$ belong to the valence band top transforming like the corresponding coordinates.

In this basis (\ref{basis}) the Kane Hamiltonian $H_{\mbox{K}}$ reads:
\onecolumngrid
\begin{equation}\label{HH0}
H_{\mbox{K}}=\left(\begin{matrix}
    0 & 0 & -Vk_+ & \sqrt{\frac23\;}Vk_z & \frac{1}{\sqrt{3}}Vk_- & 0 & -\frac{1}{\sqrt{3}}Vk_z & -\sqrt{\frac23\;}Vk_- \\
    0 & 0 & 0 & -\frac{1}{\sqrt{3}}Vk_+ & \sqrt{\frac23\;}Vk_z & Vk_- & -\sqrt{\frac23\;}Vk_+ & \frac{1}{\sqrt{3}}Vk_z \\
    -Vk_- & 0 & -E_g & 0 & 0 & 0 & 0 & 0  \\
    \sqrt{\frac23\;}Vk_z & -\frac{1}{\sqrt{3}}Vk_- & 0 & -E_g & 0 & 0 & 0 & 0 \\
    \frac{1}{\sqrt{3}}Vk_+ & \sqrt{\frac23\;}Vk_z & 0 & 0 & -E_g & 0 & 0 & 0 \\
    0 & Vk_+ & 0 & 0 & 0 & -E_g & 0 & 0  \\
    -\frac{1}{\sqrt{3}}Vk_z & -\sqrt{\frac23\;}Vk_- & 0 & 0 & 0 & 0 & -E_g-\Delta & 0 \\
    -\sqrt{\frac23\;}Vk_+ & \frac{1}{\sqrt{3}}Vk_z & 0 & 0 & 0 & 0 & 0 & -E_g-\Delta \\
\end{matrix}\right).
\end{equation}
\twocolumngrid
Here $k_\pm=(k_x\pm ik_y)/\sqrt{2}$, $V$ is an interband velocity matrix element.

Quantize the states in the $z$-direction and consider the states near the edges of the corresponding bands. The quantization can be made by the addition of the potential and the band offsets in the Hamiltonian diagonal. Omitting $k_\pm$, we have for the Hamiltonian (\ref{HH0}) projection onto these states:
\onecolumngrid
\begin{equation}\label{HH}
H_{0}=\left(\begin{matrix}
    U_c(z) & 0 & 0 & \sqrt{\frac23\;}Vk_z & 0 & 0 & -\frac{1}{\sqrt{3}}Vk_z & 0 \\
    0 &U_c(z) &0 & 0 & \sqrt{\frac23\;}Vk_z & 0 & 0 & \frac{1}{\sqrt{3}}Vk_z \\
    0 & 0 & U_v(z)-E_g & 0 & 0 & 0 & 0 & 0 \\
    \sqrt{\frac23\;}Vk_z & 0 & 0 &U_v(z) -E_g & 0 & 0 & 0 & 0 \\
    0 & \sqrt{\frac23\;}Vk_z & 0 & 0 & U_v(z)-E_g & 0 & 0 & 0 \\
    0 & 0 & 0 & 0 & 0 & U_v(z)-E_g & 0 & 0 \\
    -\frac{1}{\sqrt{3}}Vk_z & 0 & 0 & 0 & 0 & 0 & U_{so}(z)-E_g-\Delta & 0 \\
    0 & \frac{1}{\sqrt{3}}Vk_z & 0 & 0 & 0 & 0 & 0 &U_{so}(z) -E_g-\Delta \\
\end{matrix}\right).
\end{equation}
\twocolumngrid
The quantities $U_c(z)$, $U_v(z)$, and $U_{so}(z)$ include the potential and offsets between the conduction ($U_c(z)$) and valence ($U_v(z)$) band bottoms, and the SO-splitted bands $U_{so}(z)$. In this approximation the states $|3/2,\pm 3/2\rangle$ become the exact eigenstates of the Hamiltonian with energies $-E_g$.

Near the band extrema the states can be described by the band quantum numbers $\nu=(S,\sigma),\sigma=\pm 1/2$ or $\nu=(J,\sigma)$, where $J=3/2$ and $\sigma=\pm 3/2,\pm 1/2,$ or, $J=1/2,\sigma=\pm1/2$ and transversal $\phi_\nu(z)$, and longitudinal $\chi_\nu(\bm{\rho})$, $\bm{\rho}=(x,y)$ dependences. The eigenfunctions of the Hamiltonian (\ref{HH}), $\phi_\nu(z)$, have the same values for $\sigma=\pm 1/2$ states. The functions $\phi_\nu(z)$ are approximately the eigenfunctions of quadratic Hamiltonians $k_z^2/2m_\nu$.

The solutions of Shr\"odinger equation $H_0\phi(z)=E\phi(z)$ can be found by the conversion of this equation into the single-band one. The quantities $k_{\pm}V$ are small, as compared with $E_g$ and $E_g+\Delta$. Besides, we shall suppose that $k_\pm\ll k_z$, and that means that we shall excite electrons and holes near the bottoms of the quantum well subbands. Owing to this smallness, we shall construct the states from the extrema of the corresponding basic states. However, the Hamiltonians of the corresponding states should contain the in-plane momentum $k_\pm$. This is achieved by mixing the electron and hole states. This produces the single-band quadratic Hamiltonians $H_S=E_{S;n}+k^2/2\mu_e$, $H_{3/2,3/2;n}=-E_g-E_{3/2,3/2;n}-k^2/2\mu_h$, $H_{3/2,1/2;n}=-E_g-E_{3/2,1/2;n}-k^2/2\mu_{lh}$ and $H_{1/2,1/2}=-E_g-E_{1/2,1/2;n}-k^2/2m_{soh}$ for electrons, heavy holes and SO-shifted holes, respectively. These states are characterized by the effective masses
$$\frac{1}{2\mu_e}=\frac{V^2}{3}\left(\frac{2}{E_g}+\frac{1}{E_g+\Delta}\right),$$
$$\frac{1}{2\mu_{lh}}=\frac{2V^2}{3E_g},$$
$$\frac{1}{2\mu_{soh}}=\frac{1}{3}\frac{V^2}{E_g+\Delta}.$$

The heavy hole mass $\mu_h $ in this approximation occurs to be infinite. However, it is formed by the other bands and can be additionally determined.

The states can be numerated by the set of numbers, $\nu={n_b,n_z, n_r,m_{e,h}}$. The quantity $n_b$running from 1 to 8 numerates the electron band for the states $\left| S\alpha \right\rangle$, $\left| S\beta \right\rangle$, $\left| 3/2,3/2 \right\rangle$, $\left| 3/2,1/2 \right\rangle$, $\left| 3/2,-1/2 \right\rangle$, $\left| 3/2,-3/2 \right\rangle$, $\left| 1/2,1/2 \right\rangle$, $\left| 1/2,-1/2 \right\rangle$, respectively. Each state is characterized by the transversal wave function $\phi_\nu(z)$ and planar wave function $e^{im_{e,h}\varphi}J_{m_{e,h}}(kr)$.

In circular quantum dots of radius $r_0$ with infinite walls, the in-plane momentum $k_{m,n_r}$ obeys the condition $J_m(k_{m,n_r}r_0)=0$.

\section*{Optical transitions}
We shall consider a square quantum well with infinite walls. Such situation is actual in the case of an empty quantum well. The selection rules require $n_{z,h}=n_{z,e}$. The in-plane matrix elements selection rules for $a_{\pm}$ and $a_z$ do not interfere with each other, and the transition probabilities are additive.

The lowest in light frequency transitions belong to the bands $n_{z,h}=n_{z,e}=1$. The in-plane selection rules for these transitions are $m_h\to m_{e}-m\pm 1$.

The interaction with the external electromagnetic field is described by Hamiltonian (\ref{HH0}) with the replacement of $k_\pm\to a_\pm$, $k_z\to a_z$ where
$a_\pm=\frac{eV}{c\sqrt{2}}(A_x\pm iA_y)$, $a_z=\frac{e}{c}VA_z$. The quantity $a_\pm$ can be rewritten as $a_\pm=\frac{eV}{c\sqrt{2}}(A_r\pm iA_\varphi)e^{\pm i\varphi}$. In the ${\cal E}_\tau$-mode, $A_r=0$, $a_\pm=\pm\frac{eV}{c\sqrt{2}} iA_\varphi e^{\pm i\varphi}$. Then
\onecolumngrid
\begin{eqnarray}\label{etau}
&&a_\pm=\mp\frac{eVqr}{m\omega\sqrt{2}}{\cal E}_0e^{i[(m\pm 1)\varphi+qz-\omega t]}
\left\{
\begin{array}{ll}
     J_m(\kappa r), & r\leq a,\\ \nonumber
     \frac{J_m(\kappa a)}{K_m(\kappa_1a)}K_m(\kappa_1r), & r>a.
\end{array}\right.\\
&&a_z=\frac{eV}{i\omega}{\cal E}_0e^{i(m\varphi+qz-\omega t)}\left\{
\begin{array}{ll}
     J_m(\kappa r), & r\leq a,\\
     \frac{J_m(\kappa a)}{K_m(\kappa_1a)}K_m(\kappa_1r), & r>a.
\end{array}\right.
\end{eqnarray}
Similarly, for the ${\cal H}_\tau$-mode,
\begin{eqnarray}\label{htau}\nonumber
&&a_\pm=\frac{iceV}{\epsilon(r)\omega^2\sqrt{2}}{\cal H}_0\times\nonumber\\
&&e^{i[(m\pm 1)\varphi+qz-\omega t]}\left\{
\begin{array}{ll}
\left(\frac{m}{r}+\frac{q^2r}{m}\right)J_m(\kappa r)\mp\frac{\kappa }{2}[J_{m-1}(\kappa r)-J_{m+1}(\kappa r)] , & r\leq a, \\
\frac{J_m(\kappa a)}{K_m(\kappa_1a)}\left\{ \left(\frac{m}{r}+\frac{q^2r}{m}\right)K_m(\kappa_1r)\mp\frac{\kappa_1}{2}[K_{m-1}(\kappa_1r)-K_{m+1}(\kappa_1r)]\right\}, & r>a.
\end{array}\right.\nonumber\\
&&a_z=-\frac{qcerV}{\epsilon(r)\omega^2m}{\cal H}_0e^{i(m\varphi+qz-\omega t)}\left\{
\begin{array}{ll}
\frac{\kappa }{2}[J_{m-1}(\kappa r)-J_{m+1}(\kappa r)] , & r\leq a, \\
\frac{\kappa_1J_m(\kappa a)}{2K_m(\kappa_1a)} [K_{m-1}(\kappa_1r)-K_{m+1}(\kappa_1r)], & r>a.
\end{array}
\right.
\end{eqnarray}

\twocolumngrid

Owing to the smallness of $r$ corresponding to the electron wavelength, as compared with the light wavelengths, one can expand Bessel functions
$J_m(r)\sim r^{|m|}/(|m|-1)!2^{|m|}$.

The in-plane matrix elements contain, for the case $A_r=0$ the selection rules $m_e=m_h+m-1$ for transition $|3/2,3/2\rangle\to|S,1/2\rangle$ and $|3/2,-3/2\rangle\to|S,-1/2\rangle$. For $m=1$ this corresponds to the conservation of the orbital momentum projection and growth of the spin projection by 1, or growth or the orbital momentum projection of it by 2 with a spin projection drop by 1.

Note that the in-plane momenta of electrons and holes $k_e$ and $k_h$ should be close to each other due to the smallness of $\kappa$.

The absorption edge belongs to the transitions between bands $|3/2,\pm 3/2\rangle$ and $|S,\alpha\rangle$, or $|S,\beta\rangle$. The selection rules allow transitions $|3/2, 3/2\rangle\to |S\alpha\rangle$ caused by $A_-$ and $|3/2, -3/2\rangle\to |S\beta\rangle$ caused by $A_+$. These transitions may conserve or not conserve the transversal quantization due to the difference of the transversal functions, belonging to the initial and final states: $\phi_{3/2,3/2}(z)\neq \phi_{S,\alpha}(z) $. However, if the quantum well has infinite walls, the states with different transversal quantum numbers $n$ become orthogonal, and transitions $n\to n'$ become forbidden for $n\neq n'$. This selects the momentum projections $\pm 3/2\to\pm 1/2$. The in-plane selection rules yield $m_e=m_h+m$.

The matrix elements are
$$M^{(m)}_{m_h,m_e}=\int\limits_0^\infty J_{m_h}(k_hr)a_\pm(r)J_{m_e}(k_er)rdr.$$
where $a_\pm$ is determined by Eqs. (\ref{etau}) or (\ref{htau}). The quantity $M^{(m)}_{m_h,m_e}$ determines the tempo of electron momentum generation:
\begin{eqnarray}
&&g_{m;m_e}= \frac{1}{2\pi I} \int\int k_ek_hdk_edk_h|M^{(m)}_{m_h,m_e}|^2 \times\nonumber \\
&& \delta\left(\frac{k_e^2}{2\mu_e}+\frac{k_h^2}{2\mu_h}-(\omega-E_g)\right).
\end{eqnarray}

\section*{Stationary electron orbital momentum density}

The stationary distribution of orbital momentum is determined by the density matrix $\rho_{m_e,m_e'}$ in magnetic quantum numbers $m_e,m_e'$. We shall deal with the diagonal elements of $\rho_{m_e,m_e}=n_{m_e}$, neglecting the non-diagonal elements. In this case, the occupation numbers $n_{m_e}$ obey the differential equation
$$\frac{\partial}{\partial t}n_{m_e}=g_{m;m_e}-\frac{n_{m_e}}{\tau_{m_e}}.$$

The orbital-momentum relaxation time $\tau_{m_e}$ is connected with the translation momentum relaxation. In the considered case, when we do not take into account the spin-orbit interaction, the electron spin within the conduction band is a conserving quantity.

\section*{Orbital momentum fate}
The orbital momentum is not a strongly conserving quantity. In fact, the usual scattering changes the translation momentum direction, which is included into the orbital momentum expression ${\bf l}={\bf r}\times{\bf k} $. Hence, the conservation of the orbital momentum is limited by usual scattering, and, in a classical system, the relaxation rate of ${\bf l}$ should coincide with that of the translation momentum relaxation.

Here we will consider the case when the spin-orbit interaction is switched off. Below we shall discuss this issue.

The impurity scattering between the states, with quantum numbers $m_e$, $m_e'$, is determined by the probability
\begin{eqnarray}\label{W}
&&W(m_e,k_e;m_e',k_e')=\frac{4n_i}{\pi R k_e}\int \frac{d^2q}{4\pi^2}|u_{\bf q}|^2\frac{\theta(2k_e-q)}{q^2(4k_e^2-q^2)}\times\nonumber\\
&&\cos^2\left(2(m_e+m_e')\arccos\left(\frac{q}{2k_e}\right)\right)\delta\left(\frac{k_e^2}{2\mu_e}-\frac{k_e'^2}{2\mu_e}\right).
\end{eqnarray}
Here ${\bf q}'={\bf k}_e-{\bf k}_e'$ is the momentum transfer, $u({\bf q}')=\int d^2\rho u({\bf \rho})e^{i\bf q'\rho}$ is the Fourier transform of the single impurity potential and $n_i$ is the impurity concentration.

A total relaxation rate from the state $m_e,k_e$ is determined by a summation over $m_e',k_e'$,
\begin{equation}\label{taum}
1/\tau_{m_e}=\sum_{m_e',k_e'}W(m_e,k_e;m_e',k_e').
\end{equation}

The summation is limited by large values of $|m_e|$ due to the system in-plane boundary $r=R$ and the requirement that the state $m_e',k_e'$ should be fitted in $r<R$. This yields $m_e^2/R^2<2k_e^2$. Owing to large $m_e'$, one can replace the summation by integration and $\cos^2$ under the integral by 1/2. The result reads
\begin{equation}\label{taum}
\frac{1}{\tau_{m_e}}=8n_i\mu_e\int \frac{d^2q'}{(2\pi)^2}\frac{k_e^2\theta(2k_e-q')}{q^2(4k_e^2-q'^2)}|u_{\bf q'}|^2.
\end{equation}

Generally speaking, for the scattering on the neutral impurities the relaxation rate $1/\tau_{m_e}$ diverges at $q'\to 0$ and $q'\to 2k_e$. The divergence is weak (logarithmic). The divergence can be eliminated by the account of a scattering widening of states $1/\tau_p$, the phase decoherence with rate $1/\tau_\varphi$ and the interlevel distance caused by the finiteness of sample $k_e/\mu_eR$. Hence, the logarithmic factor should be written as $\ln(k_e^2\tau_\varphi/\mu_e)$, $\ln(k_e^2\tau_k/\mu_e)$, or $\ln k_eR$.

Besides, the divergence at $q'\to 0$ is eliminated if $u_{\bf q'}\to 0$. This is the case when the scattering is caused by the charged impurities screened by electrons in the quantum well and the gate electrode. In this case
$$u_{\bf q'}=\frac{2\pi e^2(1-e^{-2q'd})}{\chi (q'+2/a_B)},$$
where $d$ is the distance to the gate and $a_B$ is the Bohr radius for electrons. This yields the logarithm of an order of $\ln(k_ed)$, if $k_ea_B\gg 1$. The divergence at $q'\to 2k_e$ can be eliminated if the potential at $2k_e$ is small for some reason. Thus, instead of Eq.(\ref{taum}), we have for charged impurities
\begin{equation}\label{taum}
\frac{1}{\tau_{m_e}}=\frac{4\pi e^4n_i\mu_e}{\chi^2}\ln(4k_ed).
\end{equation}

Aside from the divergence, $\tau_{m_e}$ has the order of the usual translation momentum relaxation time $\tau_p$. This can be easily interpreted by the fact that any rotation of ${\bf k}_e$ is due to the scattering results in the change of the orbital momentum.

Probably, one should determine the relaxation rate in a different way. For example, one can determine $1/\tau_{m_e}$ via mean orbital momentum losses:
$$\frac{{m_e}}{\tau_{m_e}}=\sum_{m_e',k_e'}({m_e}-{m_e}')W({m_e},k_e;{m_e}',k_e').$$
However, the term with $m_e'$ vanishes due to the symmetry $W(m_e,k_e;m_e',k_e')=W(m_e,k_e;-m_e',k_e')$, and this determination is reduced to Eq.~\ref{taum}.

We shall assume that the holes relax much faster than electrons. In this case, we will not be interested in their momentum. The mean electron momentum projection $\langle m_e\rangle$ is determined from the kinetic equation
\begin{equation}\label{mean m}
\langle m_e\rangle=\sum\tau_{m_e}g_{m;m_e}.
\end{equation}

We performed the numerical calculations for the twisted light illumination of the GaAs-based system with energy gap $E_g=1.424$~eV. The following parameters are used: effective electron and heavy hole masses $\mu_e=0.067\mu_0$, and $\mu_h=0.45\mu_0$, where $\mu_0$ is the bare electron mass, the waveguide core radius $a=10^{-4}$~cm and the permittivities $\varepsilon=2.25$, $\varepsilon_1=1.5$. The ${\cal E}_\tau$- mode was considered only.

\begin{figure}[ht]
\centerline{\epsfysize=5cm\epsfbox{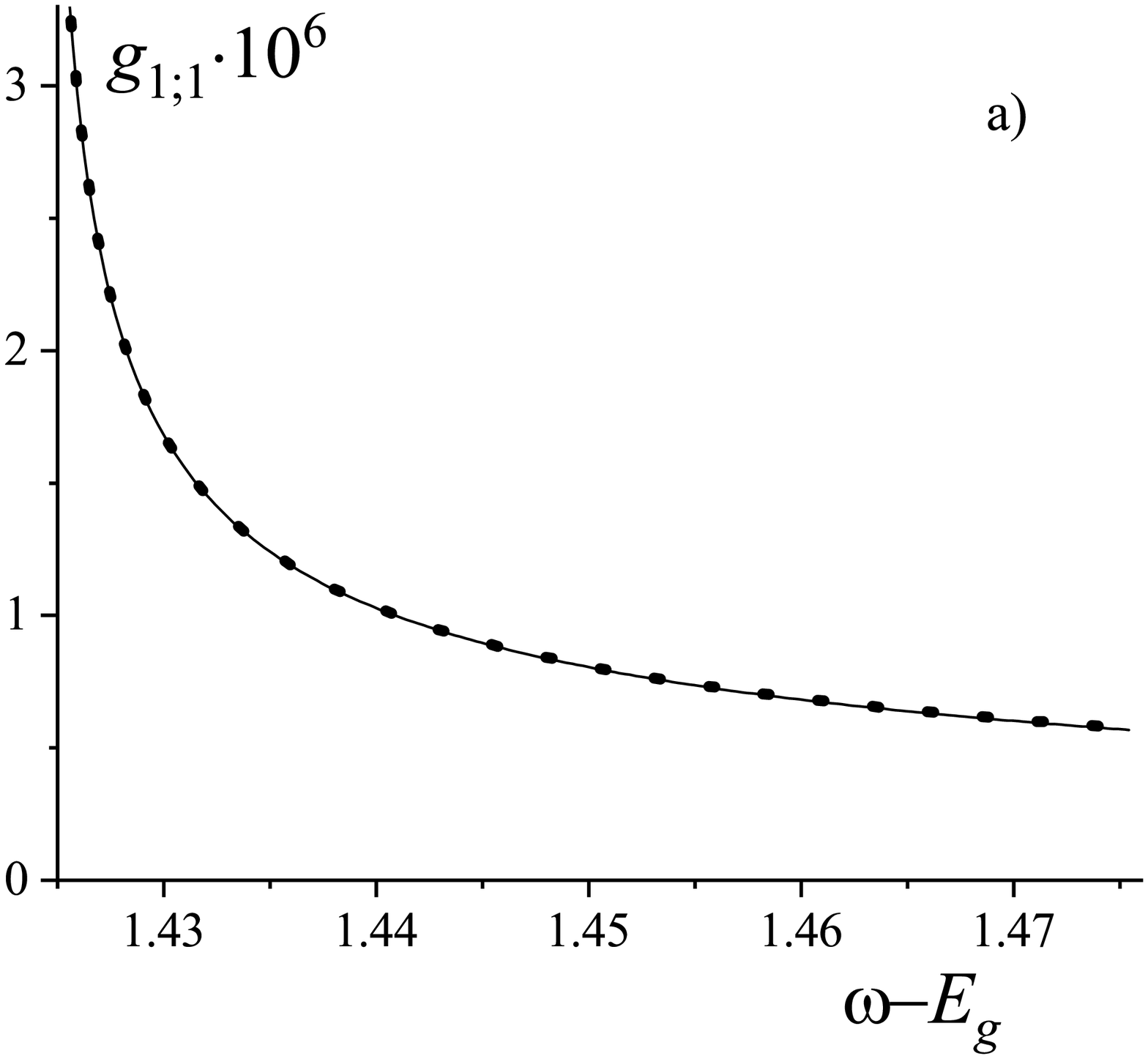}}
\centerline{\epsfysize=5cm\epsfbox{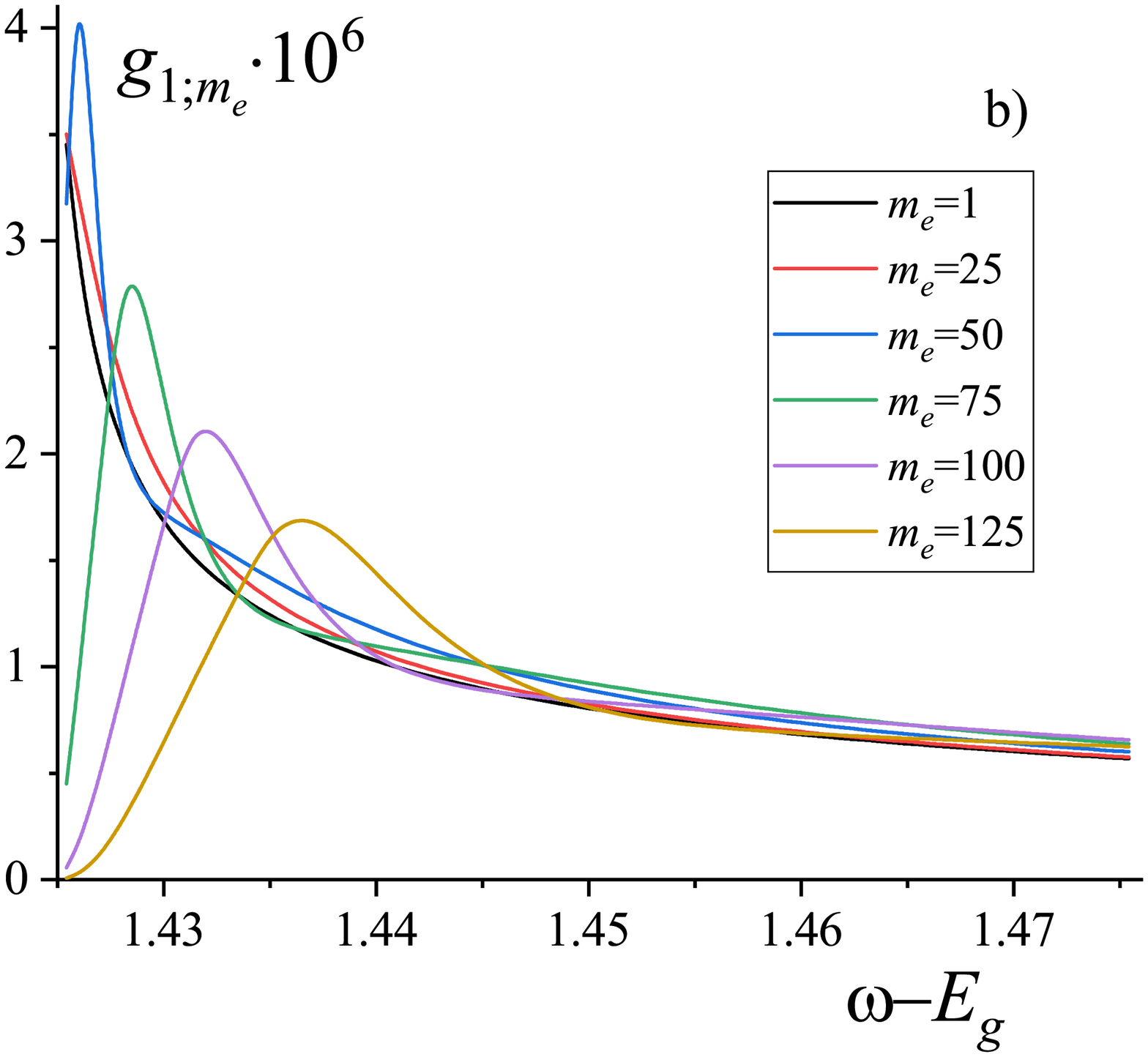}}
\caption{a) Frequency dependence of the excitation probability for $m=1,~m_e=1$ (black), as compared with the functions $0.13(\omega-E_g)^{-1/2}$ (black points). b) The same as in Figure a) for $m=1, m_e=1,25,50,75,100,125$.}\label{fig3}
\end{figure}

\begin{figure}[ht]
\centerline{\epsfysize=5cm\epsfbox{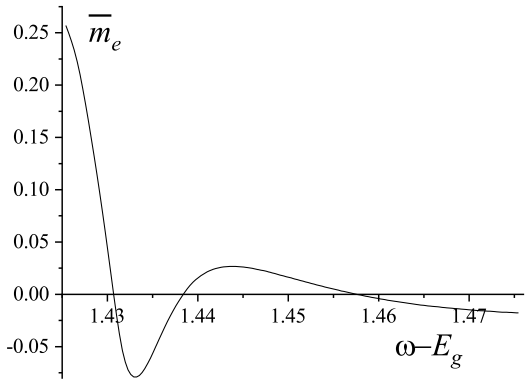}}
\caption{Frequency dependence of the mean orbital momentum projection per absorbed photon at $m=1$.}\label{fig4}
\end{figure}

The calculated mean generation rate  $m_e g_{m;m_e}$ for  fixed twisted photon and electron momenta $m=1,~m_e=1$ {\it versus} light frequency is presented in Figure~\ref{fig3}a. This dependence  follows the one-dimensional density of states $0.13(\omega-Eg)^{-1/2}$. For other electron momenta the frequency dependences deflect from this law (Figure~\ref{fig3}b). In fact, this is a consequence of the vanishing of electron Bessel functions at a small $r$, which manifests itself at a sufficiently large $m_e$ due to the slowness of function $a_\pm(r)$. The small generation rate value is due to a small absolute absorption of the 2D layer. We have also calculated the mean orbital momentum projection per absorbed photon at $m=1$,
\begin{equation}\label{memean}
 \overline{m}_e  = \frac{\sum\limits_{m_e}m_e g_{1;m_e}}{\sum\limits_{m_e}g_{1;m_e}}.
\end{equation}
(Figure~\ref{fig4}) which has the order of 1. The quantity $\overline{m}_e$ reaches its maximal value at the threshold $\omega=E_g$.

\section*{Discussion and conclusions}
Thus, we found the orbital momentum injection to the 2D electron system like a GaAs layer due to the illumination of twisted light mode from a lightguide. The twisting can be produced by the mixing light from 3 or more lightguides with phase shifts. The interband near-fundamental illumination of a semiconductor was considered. The orbital momentum generation rate was found. The orbital momentum was shown to be transferred to electrons. The relaxation rate of orbital momentum, which determines the stationary orbital momentum of 2D electrons, was found. Its rate is of the same order as that of the translation momentum order one. However, it additionally includes the logarithmic divergencies due to the decoherence processes. The estimations show that the electron orbital momentum is observable by the hot  polarized luminescence with time and space resolutions.

The orbit relaxation time due to the translation momentum non-conservation is small. Thus, the mean orbit momentum in the stationary conditions is small. However, a mean orbit momentum of hot electrons is the order of the ratio of orbit momentum and electron energy relaxation times. Usually, this ratio is not so small. Thus, the pumped orbital momentum conservation degree is expected to be large enough.

The orbital momentum is not a self-conserved quantity due to the spin-orbit interaction. However, there are ways to separate the orbital and spin momenta. This is the case when the GaAs-like semiconductor has the (1,1,1) growth direction and Rashba SIA and Dresselhaus BIA SO interaction mechanisms have similar Hamiltonians. Then one can suppress the SO-interaction \cite{Dyak-Koch,Gan}. Although Rashba SO is not a unique reason of the SO in the quantum wells, the additional Dresselhaus SO interaction can be suppressed by a suitable well orientation, by the mutual compensation of Rashba and Dresselhaus contributions, in particular, via the gate electrode-control \cite{Cart,Vur,Sun,Bal,Ye,Bier,Her,Wang}. In that case, the orbital momentum becomes conserving unless the impurity scattering plays its role. We hope that the twisted light illumination may be a possible way to the orbital momentum pumping.

Note that the orbital momentum of 2D electron gas can be observed by a hot polarized luminescence, which is a process reverse to the pumping. For example, this may be done via the same lightguide if one can detune the observation wavelength from the pumping one. The installation like that shown in Figure~\ref{fig1} can serve also as twisting-sensitive waveguide.

\section*{Acknowledgement} This research was supported by the RFBR, grant No. 20-02-00622.

\end{document}